# Investigation of Unique Total Ionizing Dose Effects in 0.2 μm Partially-Depleted Silicon-on-Insulator Technology[*]

(Has been submitted to the Chinese Physics C)

ZHANG Yan-Wei(张彦伟)[1, 2;1)], HUANG Hui-Xiang(黄辉祥)[2], BI Da-Wei(毕大炜)[2], PENG-Chao(彭超)[2], TANG Ming-Hua(唐明华)[1], ZHANG Zheng-Xuan(张正选)[2]

[1] Key Laboratory of Low Dimensional Materials and Application Technology, Xiangtan University, Ministry of Education, Xiangtan, Hunan 411105, China

[2] State Key Laboratory of Functional Materials for Informatics, Shanghai Institute of Microsystem and Information Technology, Chinese Academy of Sciences, Shanghai 200050, China

1) E-mail address: yanweizhang@mail.sim.ac.cn.

* Supported by National Science Foundation of China (61106103)

**Abstract:** The total ionizing dose (TID) irradiation effects of partially-depleted (PD) silicon-on-insulator (SOI) devices which fabricated with a commercial 0.2 μm SOI process are investigated. Experimental results show an original phenomenon that the "ON" irradiation bias configuration is the worst-case bias for both front-gate and back-gate transistor. To understand the mechanism, a charge distribution model is proposed. We think that the performance degradation of the devices is due to the radiation induced positive charge trapped in the bottom corner of shallow trench isolation (STI) oxide. In addition, comparing the irradiation responses of short and long channel devices under different drain bias, the short channel transistors show a larger degeneration of leakage current and threshold voltage. The dipole theory is introduced to explain the TID enhanced short channel effect.

## 1 Introduction

SOI technology offers many intrinsic advantages over bulk-silicon technology, including total device isolation, speed, and density, especially for the military and space applications [1]. The presence of buried oxide between top-silicon film and Si-substrate can limit the charge collection volume and improve the immunity of integrated circuits to resist the single-event upset (SEU) from energetic cosmic particles effectively. Moreover, the latch-up of devices can be eliminated drastically thanks to the complete dielectric isolation [2]. However, the TID irradiation responses of SOI transistors are more complex than bulk-silicon devices. Besides the gate and parasitic field leakage current, which are common to SOI and bulk-silicon devices, irradiation induced charge trapped in the SOI buried oxide can also affect SOI device performance [3]. With technology scaling into the submicron regime, suppressing short channel effect (SCE) has become one of the most critical issues [4, 5]. The SCE can be chiefly attributed to the DIBL effect which causes a reduction in the threshold voltage as the channel length decreases [6]. In addition, the SCE is also influenced by substrate biasing [7], buried oxide thickness, silicon-film thickness and channel doping profile [8]. In recent years, many efforts have been made to explore effective ways of improving the SCE, for example, the dual-material gate (DMG) structure [6] and the T-shaped body structure [9].

In this paper, we present the TID irradiation results of 0.2 μm gate length PD SOI NMOSFETs under three different irradiation bias configurations. The effects of radiation induced charge buildup in STI oxide and buried oxide are examined.

## 2 Experiment details

Experimental PD SOI NMOSFETs were fabricated with a 0.2 μm SOI CMOS technology. STI is introduced for isolation and the trench oxide thickness is about 400 nm. Processing was performed on a 200 mm diameter SIMBOND® wafer with 135 nm thick top silicon film and 1 μm thick buried oxide. A 3 nm thick gate oxide was grown for the test device. External body contacts were used in the T-gate devices. Irradiation experiment was performed at Xinjiang Technical Institute of Physics and Chemistry, Chinese Academy of Sciences, using a $^{60}$Co γ-rays source, typically at a dose rate of 100 rad(Si)/s. During radiation exposure, the bias conditions were consistent with the usual bias of transistors in digital circuits, i.e., the "ON" bias configuration with the gate high and the source and drain grounded, the "OFF" bias configuration with the drain high and the gate and source grounded and a common transmission gate (TG) bias configuration with the gate grounded and the source and drain high, summarized in Table1, among them, the high bias voltage $V_{DD}$=2.5 V. The devices parameters were measured prior to and after irradiation within 15 min. All the experimental curves were obtained with a Keithley 4200 semiconductor parameter analyzer. The body is grounded during measurement, as long as we do not state otherwise.

Table 1. Bias conditions during irradiation

| Bias | Drain | Source | Gate | Body | Substrate |
|---|---|---|---|---|---|
| ON | 0 | 0 | $V_{DD}$ | 0 | 0 |
| OFF | $V_{DD}$ | 0 | 0 | 0 | 0 |
| TG | $V_{DD}$ | $V_{DD}$ | 0 | 0 | 0 |

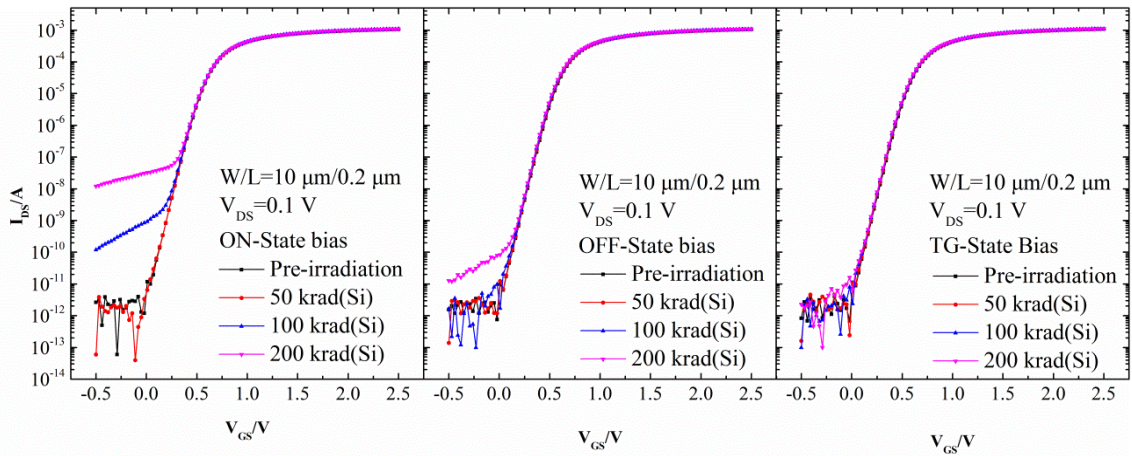

**Fig. 1.** *I-V* transfer characteristic curves for front-gate versus dose of 0.2 μm gate length PD SOI NMOSFETs under "ON", "OFF" and "TG" irradiation bias conditions.

## 3 Experimental results and discussion

3.1 Response of TID during different bias conditions

Fig. 1 shows a series of *I-V* transfer characteristic curves for front-gate transistors with *W/L*=10

μm / 0.2 μm irradiated in steps up to 200 krad(Si) under different bias conditions. The source was grounded and the drain voltage was 0.1 V during measurement. As shown on Fig. 1, for "ON" irradiation bias condition, the front-gate threshold voltage of device is almost not affected by irradiation dose. As we know, scaling of the gate oxide in advanced technology to thinner dimensions has eliminated the threshold voltage roll-off due to the irradiation induced trapped charge buildup in gate oxide [10-12]. For our transistors, the thickness of gate oxide is 3 nm, it is thin enough to avoid the direct influence of total dose degradation on the electrical characteristics of the devices. Ref. [13] indicated that charge trapped in the top corner of STI oxide induced the threshold voltage shift or sub-threshold hump effect and the increase of off-state leakage current was caused by the charge trapped in the bottom corner of STI oxide. From the above results, it can be inferred that almost no charge trapped in the top corner of STI oxide. This is because of the high electric field in the gate region drive the positive trapped charge away the top corner of STI oxide to the bottom corner. Fig. 2 shows the charge distribution model in the STI corner structure of devices, which were constructed according to the TEM picture. Positive charge trapped in the STI sidewall formed a conductive channel through which the leakage current can flow from source to drain when the parasitic transistor turned on [14]. As a consequence, the off-state leakage current increases sharply with accumulated total dose. In Fig. 1, during "ON" bias, for irradiation dose level up to and including 50 krad(Si), the leakage current does not reveal any change. After a total dose irradiation level of 100 krad(Si), the off-state current has already increased intensively. When the total dose accumulates to 200 krad(Si), a very large leakage current about $10^{-8}$ A is measured. In order to research the influence of charge trapped in buried oxide on the front-gate threshold voltage shift and leakage current, the back-gate I-V transfer characteristic curves were plotted in Fig. 3. In fact, the curves do not reflect the real back-gate transfer characteristics but the parasitic STI lateral transistor. The pre-irradiation back-gate threshold voltage for transistor can be calculated by the equation:

$$V_{T,BG} = V_{FB} + 2\Phi_B + \frac{\sqrt{2\varepsilon_s q N_A (\Phi_B)}}{C_{box}} \qquad (1)$$

where $V_{FB}$ represents the back-channel flat band voltage, $\Phi_B$ is the body potential, $\varepsilon_s$ is $SiO_2$ permittivity, $N_A$ is the back-channel doping concentration, and $C_{box}$ is the buried oxide capacitor per unit area. For convenience we adopt TCAD simulations to determine the back-transistor threshold voltage. 2D device structure and doping profile were created following the process flow. The inspection of the 0.2 μm gate length plot of $I_{DS}$-$V_{BG}$ characteristic shows that the threshold voltage of the main back-channel transistor is 56 V which is much higher than the extracted data from Fig. 3. This is consistent with our previous conclusions in Ref. [15] where the 0.13 μm PD SOI NMOSFETs were fabricated in the same foundry and confirms that parasitic STI lateral corner transistor dominates the sub-threshold behavior which we call "back-gate sub-threshold hump effect" (BGHE). According to the difference of gate length, the "TG" or "OFF" bias is the worst bias configuration for buried oxide [16]. Irradiation induced positive charge trapped in the interface of silicon and buried oxide can decrease the threshold voltage even directly inverts the channel of parasitic back-gate transistor and coupling with front-gate transistor thereby lead to deteriorative leakage current. But, this is not the case for our devices. It can be found from the back-gate characteristic curves, the "ON" bias still is the worst bias for sub-threshold region. This is due to the complete contact between STI and buried oxide. In other words, the STI oxide is the main contributor to performance degradation of transistors no matter front-gate or back-gate. In Fig. 1, these I-V characteristic curves of front-gate transistor are mainly motionless with increasing dose under "OFF" and "TG" bias in line with our discussion. From the back-gate characteristic curves we also verify the conclusions that charge trapped in the top corner of STI oxide induced the threshold voltage shift or sub-threshold hump effect and the increase of off-state leakage current is caused by the charge trapped in the bottom corner of STI oxide. Because the bottom corner of STI oxide for front-gate transistor has converted to the top corner for back-gate transistor.

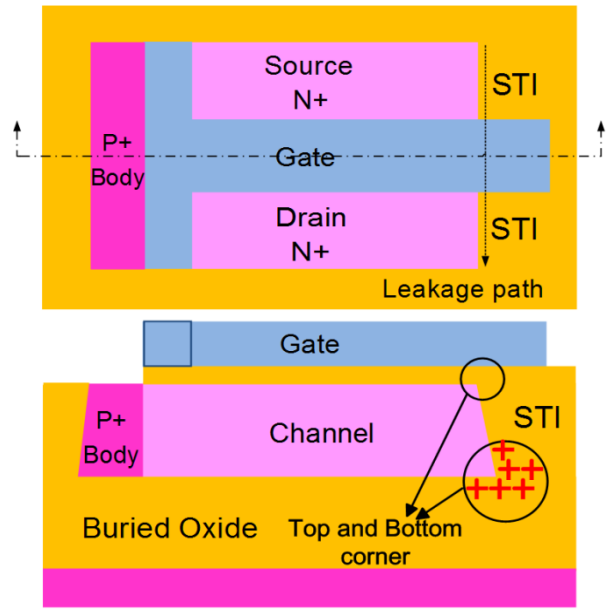

**Fig. 2.** A charge distribution model of the STI corner structure.

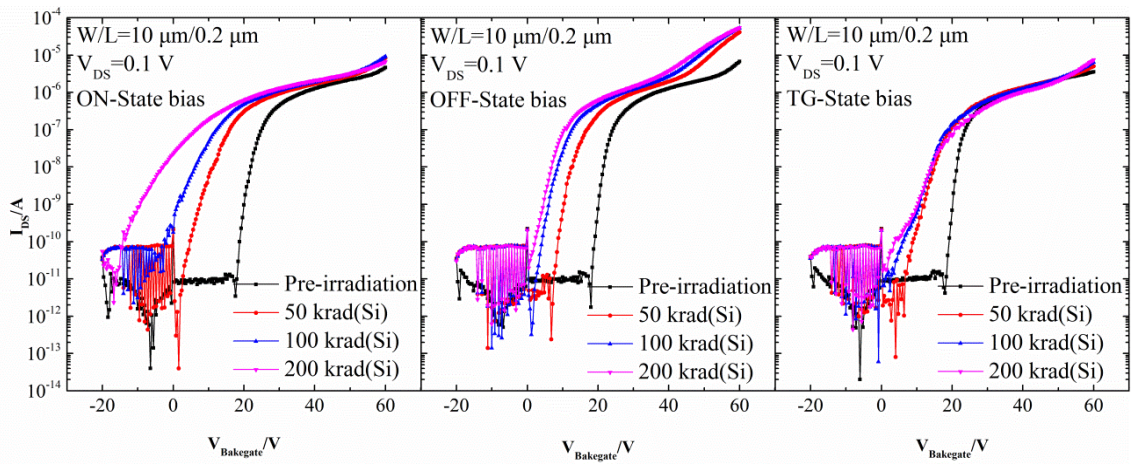

**Fig. 3.** *I-V* transfer characteristic curves for bake-gate versus dose of 0.2 μm gate length PD SOI NMOSFETs under "ON", "OFF" and "TG" irradiation bias conditions.

3.2 Response of TID during different drain voltages

Fig. 4 illustrates the *I-V* characteristic curves measured on *W/L*=10 μm /0.2 μm and *W/L*=10 μm /10 μm PD SOI NMOSFETs irradiated in steps up to 1 Mrad(Si) under different drain bias conditions. The irradiation bias configuration of transistors was "OFF" and the body was floating during measurement. As shown in Fig. 4 (a), for short channel device the threshold voltage reduces obviously when measured at a high drain voltage bias with $V_{DS}$=2.5 V compared with measured at a low drain voltage bias condition with $V_{DS}$=0.1 V. The so-called floating body effect in PD SOI

NMOSFETs is the predominant cause to lead to the threshold voltage shift. High drain voltage intensifies the impact ionization near drain region and the generated holes are collected by body. The accumulation of holes positively biases the body region then reduces threshold voltage. However, the holes generation rate due to the impact ionization decreases as gate length increases under a same value of $V_{DS}$ [17]. Therefor, for long channel device the threshold voltage reduces indistinctively. On the other hand, the short channel device measured under high drain voltage bias condition, the off-state leakage current increases signally after a total dose irradiation of 50 krad(Si). For irradiation level up to 500 krad(Si), the drain current is approximately 4 orders of magnitude higher than the corresponding pre-irradiation value. As expected, for long channel devices, whichever the drain voltage bias is, the leakage current mainly remains unchanged. The DIBL effect is used to explain this phenomenon. For short channel device, the potential barrier between source and channel region can be impacted intensively by drain electric field. Upon application of a high drain voltage bias condition, the barrier height for channel barriers at the edge of the source reduces due to the influence of high drain electric field. As a consequence, the number of carriers that injected into the channel region from the source increased and that leaded to the increase of off-state current. In Fig. 4 (a), the leakage current increases with the increase of TID. However, from the above discussion we know the STI sidewall parasitic transistor is unopened during "OFF" irradiation bias condition. In order to understand the irradiation induced degeneration in off-state current of short channel device, the dipole theory is introduced to explain this phenomenon. The 3D-DIBL effect of MOSFET isolated with STI and explanation with dipole theory were discussed by C. H. Wang and P. F. Zhang [18]. Before TID irradiation, as shown in Fig. 5, the charge imaging effect takes place from the drain charge to the image charge in the gate electrode over the STI region. This sidewall charge imaging effect physically reduces the drain to source coupling field. This "proximity effect" not only reduces the total field lines penetrating into the source region, but also causes the field lines near the width center to spread outwards toward the channel edge, and hence further weakens the drain-to-source coupling. After TID irradiation, the holes were captured by the hole-traps in the STI oxide and buried oxide. The trapped positive charge will influence the distribution of electric field lines. As shown in Fig. 6, on one hand, the fringing field from drain to gate will be cut off by charge trapped in the STI. The decrease of drain to gate coupling will enhance the drain to source coupling and enhances the DIBL effect. The electric field lines near the width center become more compact and improve drain to source coupling. Ref. [19] discusses a similar result in bulk-silicon MOSFET. On the other hand, the charge trapped in the buried oxide will enhance the back interface to source bottom coupling and lower the potential barrier of back channel from drain to source. The coupling effect will increase the likelihood of back channel conduction. Fortunately, for our PD SOI devices, it will not affect the front channel. However, it will be a significative issue to study the DIBL effect of back channel and coupling effect between front channel and back channel for FD SOI transistors.

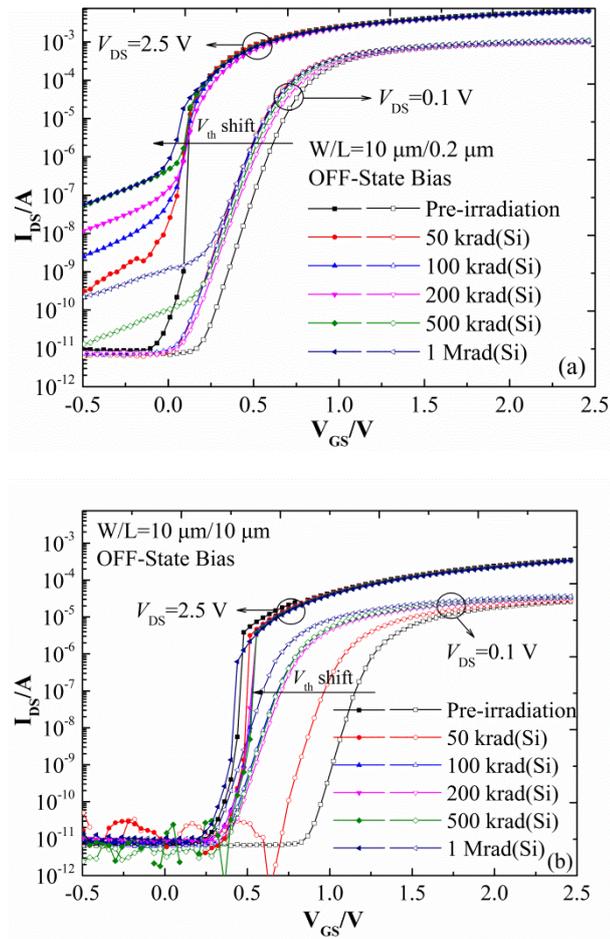

**Fig. 4.** *I-V* transfer characteristic curves versus dose of (a) W/L=10 μm /0.2 μm and (b) W/L=10 μm /10 μm PD SOI NMOSFETs with body floating at a high drain voltage bias (solid symbols) and the low drain voltage bias (open symbols) conditions. The irradiation bias configuration was "OFF".

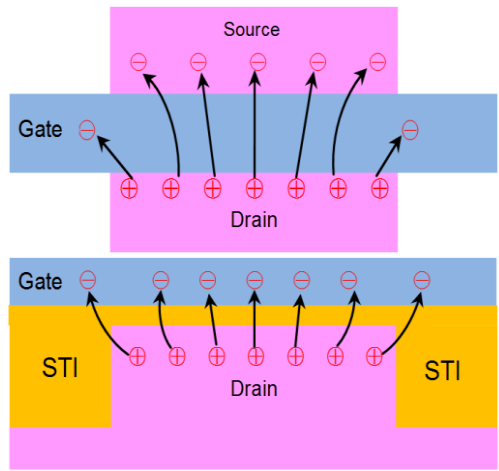

**Fig. 5.** A simple dipole representation of the charge imaging effect and the resulted field pattern for 3-D DIBL.

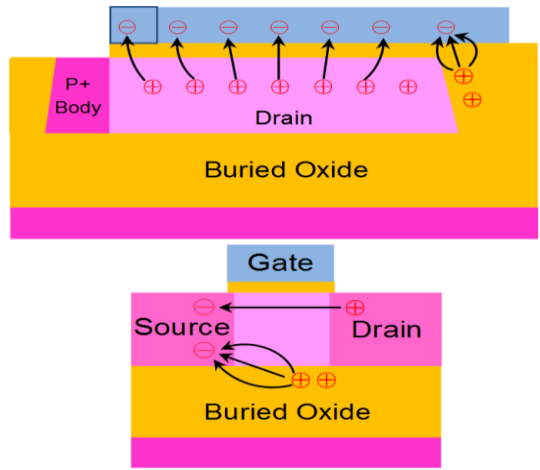

**Fig. 6.** The coupling effects from drain to gate and from buried oxide to bottom of source after TID irradiation.

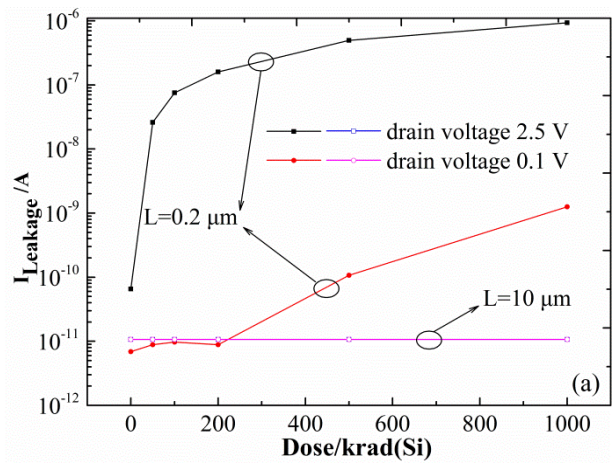

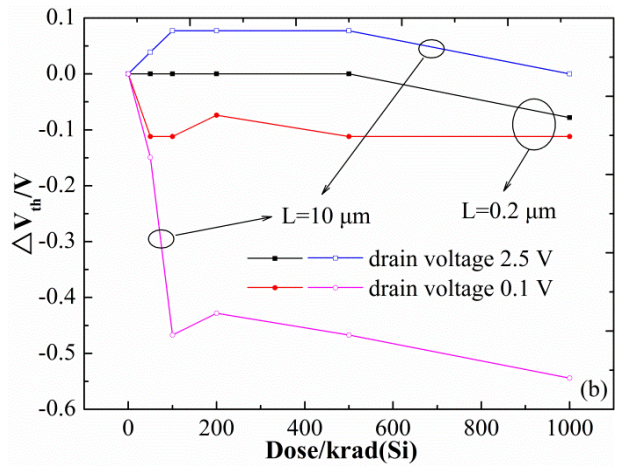

**Fig. 7.** (a) Off-state leakage current and (b) Front-gate threshold voltage shift versus TID of different gate length PD SOI NMOSFET measured under different drain bias conditions.

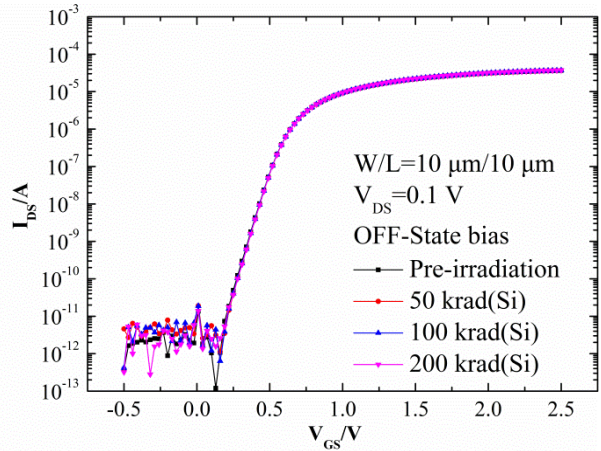

**Fig. 8.** *I-V* transfer characteristic curve versus dose of 10 μm gate length PD SOI NMOSFET with external body contact.

Furthermore, in Fig. 4 (b), the threshold voltage shift is obvious for the low drain voltage bias. This situation does not appear in the *I-V* curves measured on 0.2 μm gate length device. At high drain voltage bias condition, the sub-threshold slope is nearly infinite. The floating body effect is partly responsible for the threshold voltage shift and steeper sub-threshold slope. We extract the off-state leakage current and threshold voltage shift from the *I-V* curves to plot the response versus the TID under different drain bias conditions. As shown in Fig. 7, it is clear that the largest leakage current occurs on short channel 0.2 μm gate length device under high drain voltage bias. However, the threshold voltage shift is remarkable for 10 μm gate length device at the low drain bias condition. This can be explained by the fact that the neutral body region is reduced with the channel length decreasing. When contact body to the ground, the threshold voltage shift disappeared, as shown in Fig. 8.

## 4 Conclusions

We have analyzed the TID response of short channel 0.2 μm gate length PD SOI NMOSFETs under different irradiation bias conditions and discussed the principal elements that lead to the sub-threshold leakage current. The results show that irradiation induced trapped charge in the bottom corner of STI oxide is the decisive factor which is responsible for front gate off-state leakage current. In addition, we compared the electrical properties of different gate length transistors versus TID and studied the influence of DIBL effect to leakage current at high drain voltage bias condition. The dipole theory indicates that positive charge trapped in the STI oxide induced the enhanced DIBL effect, as a consequence, the off-state leakage current of short channel devices are impacted by irradiation more susceptibly.

## Acknowledgments

The authors are indebted to the Xinjiang Technical Institute of Physics & Chemistry, Chinese Academy of Sciences for the radiation experiment and useful discussions.